\documentclass[longbibliography, aps, prx, preprint]{revtex4-2}
\usepackage[utf8]{inputenc}
\usepackage{amsmath}
\usepackage{color}
\usepackage{hyperref}
\usepackage{graphicx}
\usepackage{setspace}
\usepackage{tikz}
\usepackage{multirow}
\usepackage{braket}

\begin{document}

\title{Strong electron-phonon coupling and carrier self-trapping in Sb$_2$S$_3$}
\author{Yun Liu}
\email{liu\_yun@ihpc.a-star.edu.sg}
\affiliation{Institute of High Performance Computing (IHPC), Agency for Science, Technology and Research (A*STAR), 1 Fusionopolis Way, \#16-16 Connexis, Singapore 138632, Republic of Singapore}
\affiliation{Cavendish Laboratory, University of Cambridge, Cambridge CB3 0HE, United Kingdom}
\author{Julia Wiktor}
\affiliation{Department of Physics, Chalmers University of Technology, SE-412 96 Gothenburg, Sweden}
\author{Bartomeu Monserrat}
\email{bm418@cam.ac.uk}
\affiliation{Cavendish Laboratory, University of Cambridge, Cambridge CB3 0HE, United Kingdom}
\affiliation{Department of Materials Science and Metallurgy, University of Cambridge, Cambridge CB3 0FS, United Kingdom}

\begin{abstract}
Antimony sulphide (Sb$_2$S$_3$) is an Earth-abundant and non-toxic material that is under investigation for solar energy conversion applications. However, it still suffers from poor power conversion efficiency and a large open circuit voltage loss that have usually been attributed to point or interfacial defects and trap states. 
More recently, there has been some discussion in the literature about the role of carrier trapping in the optoelectronic properties of Sb$_2$S$_3$, with some reporting self-trapped exciton (STE) as the microscopic origin for the performance loss, while others have found no evidence of carrier trapping with only large polaron existing in Sb$_2$S$_3$. 
By using first-principles methods, we demonstrate that Sb$_2$S$_3$ exhibits strong electron-phonon coupling, a prerequisite for carrier self-trapping in semiconductors, which results in a large renormalization of $200$\,meV of the absorption edge when temperature increases from $10$\,K to $300$\,K. 
When two electrons or holes are added to the system, bipolarons are observed with localized charge density accompanying significant lattice distortion with the formation of Sb and S dimers.
When the bipolarons are placed near each other, a bi-STE with formation energy per exciton of $-700$\,meV is observed, in general agreement with the experimentally measured Stokes shift.
Our results reconcile some of the controversy in the literature regarding the existence of carrier trapping in Sb$_2$S$_3$, and demonstrate the importance of systematically investigating electron-phonon coupling and polaron and STE formation in the antimony chalcogenide family of semiconductors for optoelectronic applications.
\end{abstract}

\maketitle

\section*{Introduction}
Photovoltaic (PV) solar cells are one of the key technologies for realizing a decarbonized economy as the Sun is an inexhaustible and clean energy source. Mainstream solar panels have been mainly based on crystalline silicon, which offers high power conversion efficiencies (PCE) at over 25$\%$ and its cost has decreased substantially over the years\cite{saga_advances_2010}. While other emerging materials such as organic-inorganic hybrid perovskites and thin film technologies such as CIGS and CdTe are making rapid improvements in PCE, they still face stability, toxicity, and material scarcity issues\cite{jean_pathways_2015}. To further increase the PCE and lower the cost of PV generated electricity, tandem solar cells show great potential as they can break the Shockley-Queisser limit of single junction solar cells\cite{ehrler_photovoltaics_2020}. The widely used silicon PV has a bandgap of around $1.1$\,eV and is an ideal material for the bottom cell to absorb the lower energy part of the solar spectrum. The search for top cell materials compatible with crystalline silicon is an active area of research for the scientific and engineering communities, with candidates ranging from III-V semiconductors to perovskites\cite{li_perovskite_2020}.

Among the many novel material candidates, the metal chalcogenide family has received a lot of attention due to their Earth-abundant and low-toxicity elements\cite{kim_highly_2014, wang_stable_2017, WANG2018713, kondrotas_sb2s3_2018}. They also possess desirable band gaps and relatively benign synthesis conditions. In particular, antimony sulphide (Sb$_2$S$_3$) has a high absorption coefficient in the visible region and a band gap of $1.7$\,eV that is ideal for the top subcell in a Si-based tandem solar cell. Despite these promising traits, the record PCE of Sb$_2$S$_3$ is only about 7.5$\%$\cite{choi_highly_2014}, far from the minimum $18$\% needed for an efficient top cell\cite{ROSSRUCKER2022178}. This is due to the fact that Sb$_2$S$_3$ suffers from high open circuit voltage ($V_{\mathrm{oc}}$) losses, even though the internal quantum efficiency is near unity and the fill factor is up to $70$\%. Irrespective of fabrication methods, the $V_{\mathrm{oc}}$ is only about $0.7$\,eV, half the theoretical maximum allowed by its band gap. This large $V_{\mathrm{oc}}$ loss has generally been ascribed to the presence of localized point defects such as sulphur vacancies or interfacial defects between Sb$_2$S$_3$ and the carrier transport layers\cite{doi:10.1021/jp4072394, cai_extrinsic_2020,  maiti_sulfur-vacancy_2020, lian_revealing_2021}. Such trap states in the band gap can act as non-radiative recombination centres to reduce photocarrier populations\cite{wang2023fourelectron}. Defects can also reduce the quasi-Fermi level splitting range under illumination and lead to lower $V_{\mathrm{oc}}$ and poor device performance.

Some recent reports have attributed the $V_{\mathrm{oc}}$ loss in metal chalcogenides to intrinsic carrier self-trapping\cite{yang_ultrafast_2019, PhysRevB.55.5799, https://doi.org/10.1002/advs.202202154}.
In Sb$_2$S$_3$, the role of extrinsic defects was excluded by the observation of a few picosecond carrier trapping without saturation at high carrier density of $10^{20}$ \,cm$^{-3}$ and the polarized nature of trap emission from single crystals\cite{yang_ultrafast_2019}. 
In Sb$_2$Se$_3$, lattice anharmonicity was observed with a 20\,ps barrierless intrinsic self-trapping with associated polaronic lattice distortion\cite{https://doi.org/10.1002/advs.202202154}.
On the other hand, a first-principles study found that polarons in these systems have rather large radii extending over several unit cells and moderate Fr\"{o}hlich coupling constants\cite{doi:10.1021/acsenergylett.2c01464}.
Therefore, the debate on the role of small localized polaron and carrier trapping in Sb$_2$S$_3$ remains open.

A prerequisite for the formation of polarons is strong coupling between the carriers and the lattice. Experimentally, the importance of electron-phonon coupling in Sb$_2$S$_3$ has been studied by Chong and co-workers, who observed coherent phonon generation in pump-probe experiments, and assigned it to the $B_{3g}$ longitudinal optical phonon mode at 65 cm$^{-1}$.
It was also reported that a $A_g$ optical phonon mode at 194\,cm$^{-1}$ is responsible for the excited state relaxation in Sb$_2$Se$_3$\cite{https://doi.org/10.1002/advs.202202154}. The electronic structure and band gaps of Sb$_2$S$_3$ were also calculated at various levels of theory\cite{caracas_first_principles_2005, ben_nasr_electronic_2011, filip_gw_2013}, and, separately, the phonon dispersion and anisotropic thermal expansion \cite{liu_first_principles_2014, gan_large_2015}. However, a full microscopic characterization of electron-phonon coupling is still missing.

In this work, we perform a systematic first-principles study of electron-phonon coupling and polarons in Sb$_2$S$_3$. We reveal the presence of strong electron-phonon coupling, leading to a large absorption edge renormalization of $200$\,meV when temperature increases from $10$\,K to $300$\,K.
We find that there are negligible structural distortions when an electron is added or removed from the supercell with the charge density remaining delocalized across the system. 
In the presence of two excess electrons or holes per supercell, corresponding to a carrier density of $10^{20}$ \,cm$^{-3}$, we observe bipolarons associated with the formation of antimony and sulphur dimers, respectively.
When the electron and hole bipolarons are placed next to each other, a bi-self-trapped-exciton (bi-STE) encompassing two neighbouring STE, is observed with a formation energy of $-700$\,meV per exciton.
Our results contribute to the debate regarding the existence and role of polarons and STE, and highlight the complex carrier self-trapping properties in metal chalcogenide systems mediated by strong electron-phonon coupling.

\begin{figure}[t]
 \includegraphics[scale=0.08]{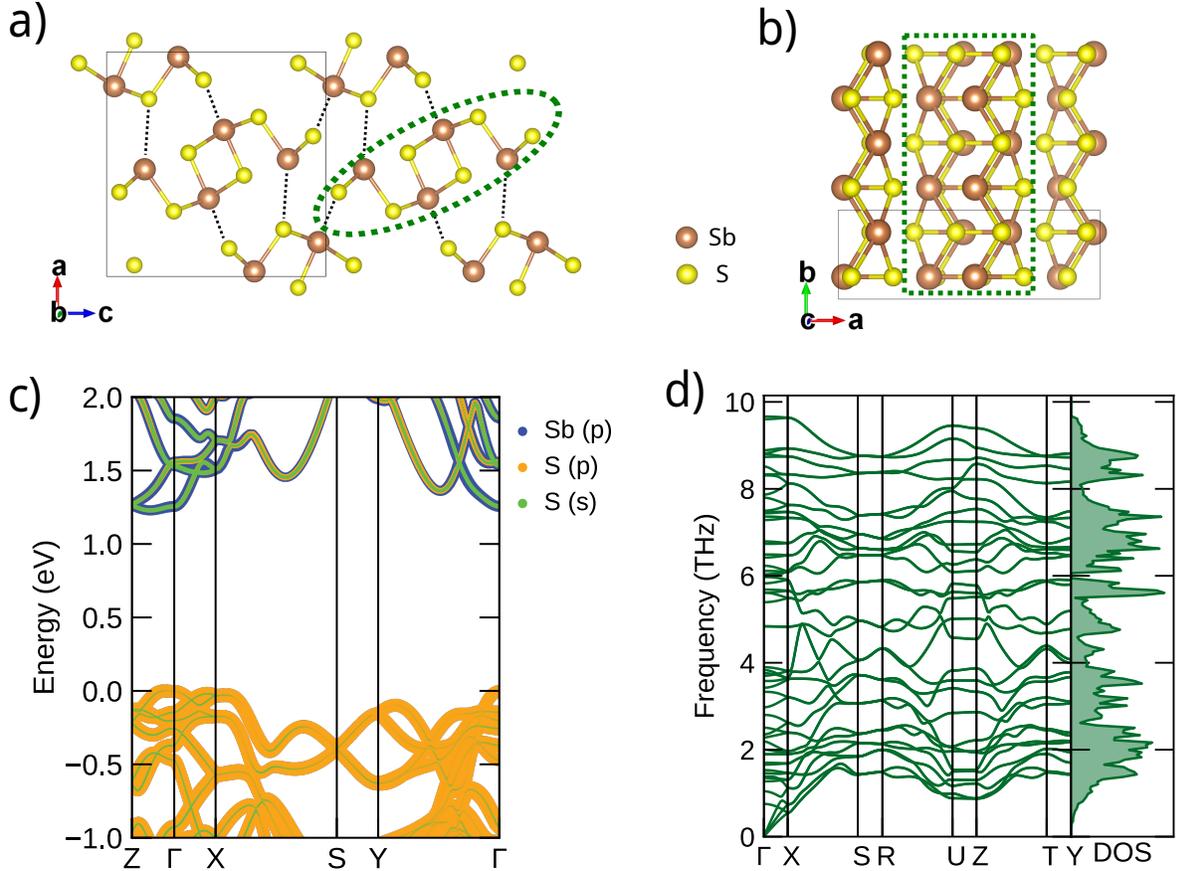}
 \centering
 \caption{(a) Sb$_2$S$_3$ crystal structure viewed from the $[010]$ axis, with the unit cell enclosed by the black box. Brown and yellow spheres represent Sb and S atoms, respectively. One [Sb$_4$S$_6$] ribbon is encircled in the green dashed line. The intra-ribbon bonds are indicated by the solid brown-yellow lines, and the inter-ribbon van der Waals interactions by black dashed lines. (b) Sb$_2$S$_3$ crystal structure viewed from the $[001]$ axis, with the ribbon enclosed in a green dashed line box. (c) The projected band structures along high symmetry lines of the Brillouin zone calculated at the optB86b level, with the orbital contributions drawn as a series of stacked circles. (d) The phonon dispersion of Sb$_2$S$_3$ along high symmetry lines of the Brillouin zone with the total phonon density of states on the right panel.}
 \label{fig_lattice}
\end{figure}

\section*{Results and Discussions}
\subsection{Equilibrium properties}
\begin{table}
\begin{tabular}{ c| c c c |c c }
\hline \hline
 & \multicolumn{3}{c|}{Lattice parameter (\AA)} & &  \\
 & $a$ & $b$ & $c$ & $E_g^{\mathrm{direct}}$ & $E_g^{\mathrm{indirect}}$ \\
 \hline
 optB86b & 11.324 & 3.865 & 11.053 & 1.241 & 1.228 \\
 SCAN+rVV10 & 11.315 & 3.843 & 11.085 & 1.362 & 1.362 \\
 SCAN & 11.672 & 3.847 & 11.254 & 1.382 & 1.382 \\
 PBE & 12.170 & 3.870 & 11.228 & 1.259 & 1.241 \\
 PBEsol & 11.267 & 3.829 & 10.908 & 1.307 & 1.287 \\
 HSE06 & 12.081 & 3.802 & 11.389 & 1.745 & 1.740 \\
 \hline
 Experiment & 11.311 & 3.836 & 11.229 & 1.7 & \\
\hline \hline
 
\end{tabular}
\caption{The lattice parameters and band gaps of Sb$_2$S$_3$ calculated using different DFT exchange-correlation functionals and compared with experimental values.\cite{bayliss1972}}
\label{table_functional}
\end{table}

The orthorhombic phase of Sb$_2$S$_3$ belongs to the space group \textit{Pbnm} with $20$ atoms per unit cell. Its crystal structure is highly anisotropic with covalently bonded 1D ribbons of Sb$_4$S$_6$ along the $[010]$ or $b$ direction (Figure \ref{fig_lattice}(a)(b)). These ribbons are in turn weakly bonded in a zigzag fashion in the $(010)$ plane by van der Waals interactions. Due to the presence of van der Waals interactions, we test the nonlocal vdW-DF functional optB86b and SCAN+rVV10\cite{klimes_van_2011, peng_versatile_2016} against some commonly used semi-local, metaGGA, and hybrid functionals\cite{perdew_generalized_1996, sun_accurate_2016, perdew_restoring_2008, heyd_hybrid_2003} (details in the Methods section). While most functionals are able to reproduce the $b$ lattice parameter accurately, the vdW functional performs the best at simultaneously reproducing $a$ and $c$ accurately due to its better performance at capturing the van der Waals interactions in the $(010)$ plane (see Table \ref{table_functional}). There are little differences in the calculated lattice parameter between optB86b and SCAN+rVV10 functionals.

We then compute the orbital-projected band structure which is plotted in Figure \ref{fig_lattice}(c). The valence band maximum (VBM) consists of mainly S $3p$ orbitals and the conduction band minimum (CBM) is dominated by Sb-S bonds made up of Sb $5p$ and S $3s$ orbitals. The position of the VBM and CBM are located slightly away from $\Gamma$ at $(0,0,0.103)$ and $(0,0,0.282)$, respectively. This means that Sb$_2$S$_3$ is an indirect bandgap semiconductor, which has been reported by previous theoretical studies\cite{filip_gw_2013}, and experimentally observed in low temperature optical measurements by Fujita and co-workers \cite{doi:10.1143/JPSJ.56.3734}.

The vdW functional suffers from the same self-interaction error as other semi-local and metaGGA functionals and underestimates the band gap to be around $1.24$\,eV(Table \ref{table_functional}). Using the hybrid functional HSE06 leads to a direct band gap of $1.75$\,eV that is close to the experimental value. Our results are in general agreement with previously calculated band gaps in the range $1.2$-$1.7$\,eV at different levels of theory\cite{filip_gw_2013, ben_nasr_electronic_2011, caracas_first_principles_2005}. Due to the small difference between direct and indirect band gaps, $\Delta(E_g^{\mathrm{direct}}-E_g^{\mathrm{indirect}}) < 20$meV, Sb$_2$S$_3$ is often treated as an effective direct band gap semiconductor. While hybrid functionals can more accurately reproduce the band gap, the computational cost is significantly larger than that of metaGGA functional like SCAN+rVV10, which in turns has higher computational cost than optB86b. We therefore use the optB86b functional for the electron-phonon calculations in this study and apply a HSE06-derived scissor shift to the band gaps when necessary. The computed phonon dispersion and density of states (DOS) is shown in Figure \ref{fig_lattice}(d), which exhibits no imaginary modes\cite{liu_first_principles_2014} indicating dynamical stability.

For polaron calculations, hybrid functionals are necessary to describe the charge localization \cite{ouhbi_polaron_2021, janotti_vacancies_2014, spreafico_nature_2014}. To determine the fraction of Fock exchange needed to cancel the self-interaction error, we verify the fulfillment of the Koopmans' condition of the screened hybrid functional. This is done by calculating the occupied and unoccupied single particle energy levels related to the $+/0$ transition of an unrelaxed sulphur vacancy, which lies within the semiconductor gap of Sb$_2$S$_3$\cite{https://doi.org/10.1002/solr.201900503}. We vary the amount of exact Fock exchange while keeping the screening parameter constant at $0.2$.
Figure S1 shows the computed band edges and the energy levels for the sulphur vacancy transitions. We note that the single-particle energy levels shown in the figure include finite-size corrections\cite{PhysRevLett.102.016402,falletta2020finite}. The crossing between the levels calculated in the $0$ and $+1$ charge of the supercell corresponds to the value of exact exchange for which the Koopmans’ condition is satisfied. As this crossing value of $0.24$ is very close to the default value of $0.25$ in HSE06, we use the default $0.25$ Fock exchange for the subsequent polaron calculations.

\subsection{Temperature dependent structural and optical properties}
\begin{figure}[t]
 \includegraphics[scale=0.38]{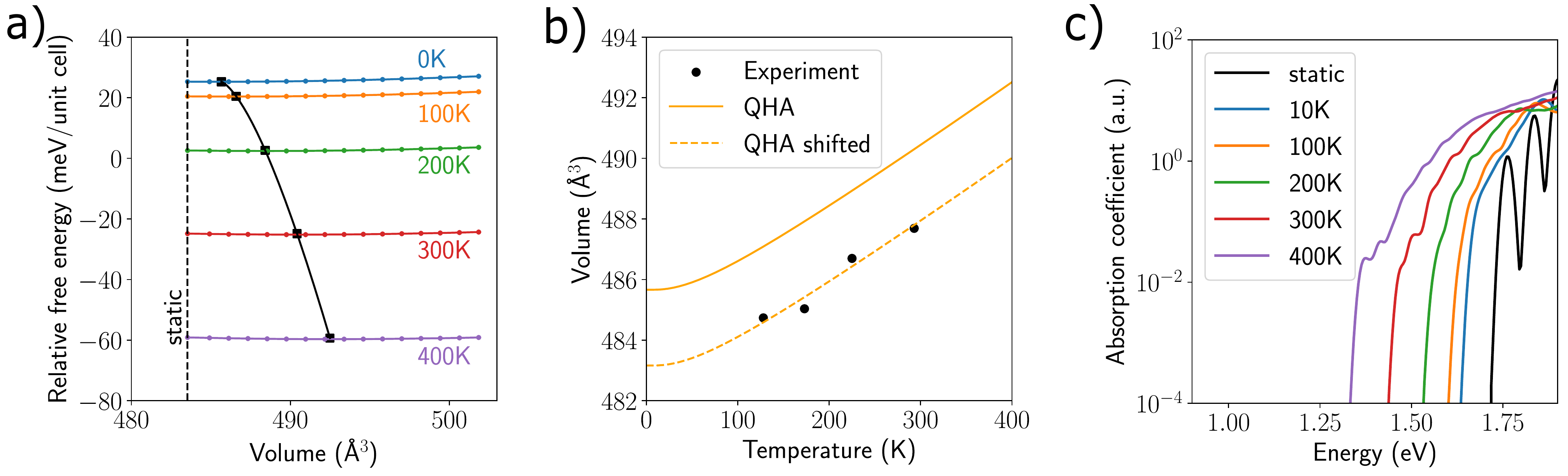}
 \centering
 \caption{(a) Relative Helmholtz free energy as a function of the unit cell volume for temperatures between $0$ and $400$\,K. The black dashed vertical line indicates the volume at the static DFT level, and the black squares indicate the minima of fitted free-energy curves at each given temperature with the Rose-Vinet equation of state. (b) The unit cell volume as a function of temperature as calculated by the QHA. The QHA data are also shifted by $-2.5$\,\AA$^3$ to guide the comparison with experimental volumes\cite{kyono_low-temperature_2002}. (c) Absorption coefficient at temperatures from 10K to 400K including phonon-assisted processes. A scissor operator of $0.5$\,eV is applied to align the static DFT bandgap to that calculated at the HSE06 level.}
 \label{fig_qha}
\end{figure}

\begin{table}
\begin{tabular}{c c c}
\hline \hline
 Material & $\Delta E_g$ (meV) & References\\
 \hline
 Sb$_2$S$_3$ & 200 & This study \\
 Bi$_2$S$_3$ & 159 & \cite{linhart_nesting-like_2021}  \\
 GaAs & 96 & \cite{doi:10.1063/1.1368156} \\
 GaSb & 85 & \cite{doi:10.1063/1.1368156} \\
 MoS$_2$ & 95 & \cite{Sigiro_2015} \\
 WSe$_2$ & 66 & \cite{C5NR01536G} \\
\hline \hline
 
\end{tabular}
\caption{The change in bandgap for selected semiconductors in the temperature range from 10 to 300K.}
\label{table_bandgap}
\end{table}

The first effect we consider for the description of the finite temperature optoelectronic properties of Sb$_2$S$_3$ is the role of thermal expansion. Using the quasi-harmonic approximation (QHA)\cite{togo_first-principles_2010}, we consider primitive cell volumes ranging from $483.5$\,\AA$^3$ to about $502$\,\AA$^3$ in $15$ equidistant steps. As the unit cell of Sb$_2$S$_3$ is orthorhombic, the lattice parameters and atomic positions at each volumetric step are relaxed while keeping the volume constant. Then the Helmholtz free energies of the relaxed structures are calculated within the harmonic approximation to the lattice dynamics. Figure \ref{fig_qha}(a) depicts the Helmholtz free energy relative to the static lattice energy as a function of lattice volume for temperatures ranging from $0$ to $400$\,K. The minimum of each fitted free-energy curve gives the quasi-harmonic volume at the corresponding temperature. The zero-point quantum motion contributes a volume increase of about $2$\,\AA$^3$, and thermal expansion increases the volume by an additional $7.5$\,\AA$^3$ in the studied temperature range. 

Figure \ref{fig_qha}(b) shows the comparison between the experimental temperature dependent volume \cite{kyono_low-temperature_2002} and the QHA results. The raw DFT calculations overestimate the volume by a constant $2.5$\,\AA$^3$ from $128$\,K to $293$\,K where experimental data are available, which is mainly due to the accuracy limits of the optB86b functional. There is remarkable agreement between theory and experiment if this systematic error is corrected, revealing that thermal expansion is correctly captured by our model.

Both $E_g^{\mathrm{direct}}$ and $E_g^{\mathrm{indirect}}$ change by about $9$\,meV when considering the influence of thermal expansion from $0$\,K to $400$\,K. This change is negligible compared to the absorption edge renormalization induced by electron-phonon coupling to be discussed next. Therefore, we will ignore the effects of thermal expansion in the rest of this work. 

Figure \ref{fig_qha}(c) shows the absorption spectrum of Sb$_2$S$_3$ in logarithmic scale calculated with the optB86b functional and a HSE06-derived scissor correction. The static absorption spectrum shows sharp features due to the relatively small smearing parameter of $15$\,meV used, which is nonetheless necessary to accurately locate the absorption onset. As the temperature increases, the optical absorption spectra are smoothed out by the continuous spectrum of allowed electronic states whose energies are renormalized by the inclusion of electron-phonon coupling effects (see Methods section for details). The absorption onset at $10$\,K is about $100$\,meV below that of the static level, indicating the importance of zero-point motion. Increasing temperature leads to a further red-shift of the absorption onset, a result that is consistent with recent experimental measurements of Bi$_2$S$_3$ in the chalcogenide family\cite{linhart_nesting-like_2021}. The calculated absorption onset changes by around $200$\,meV from $10$\,K to $300$\,K, somewhat larger than that of $159$\,meV reported for Bi$_2$S$_3$, but the difference likely arises from the mass difference between Bi and Sb\cite{doi:10.1021/acs.jpcc.1c00861}. Absorption onset redshifts of about $100$\,meV have been observed in other conventional III-V or van der Waals semiconductors (see Table \ref{table_bandgap}).

The large redshift in the temperature dependent absorption onset shows that Sb$_2$S$_3$ exhibits large electron-phonon coupling. Such strong coupling between electrons and phonons provides the necessary conditions for the possible generation of small localized polarons and STE which we discuss next.

\subsection{Carrier trapping}
\begin{figure}[t]
 \includegraphics[scale=0.25]{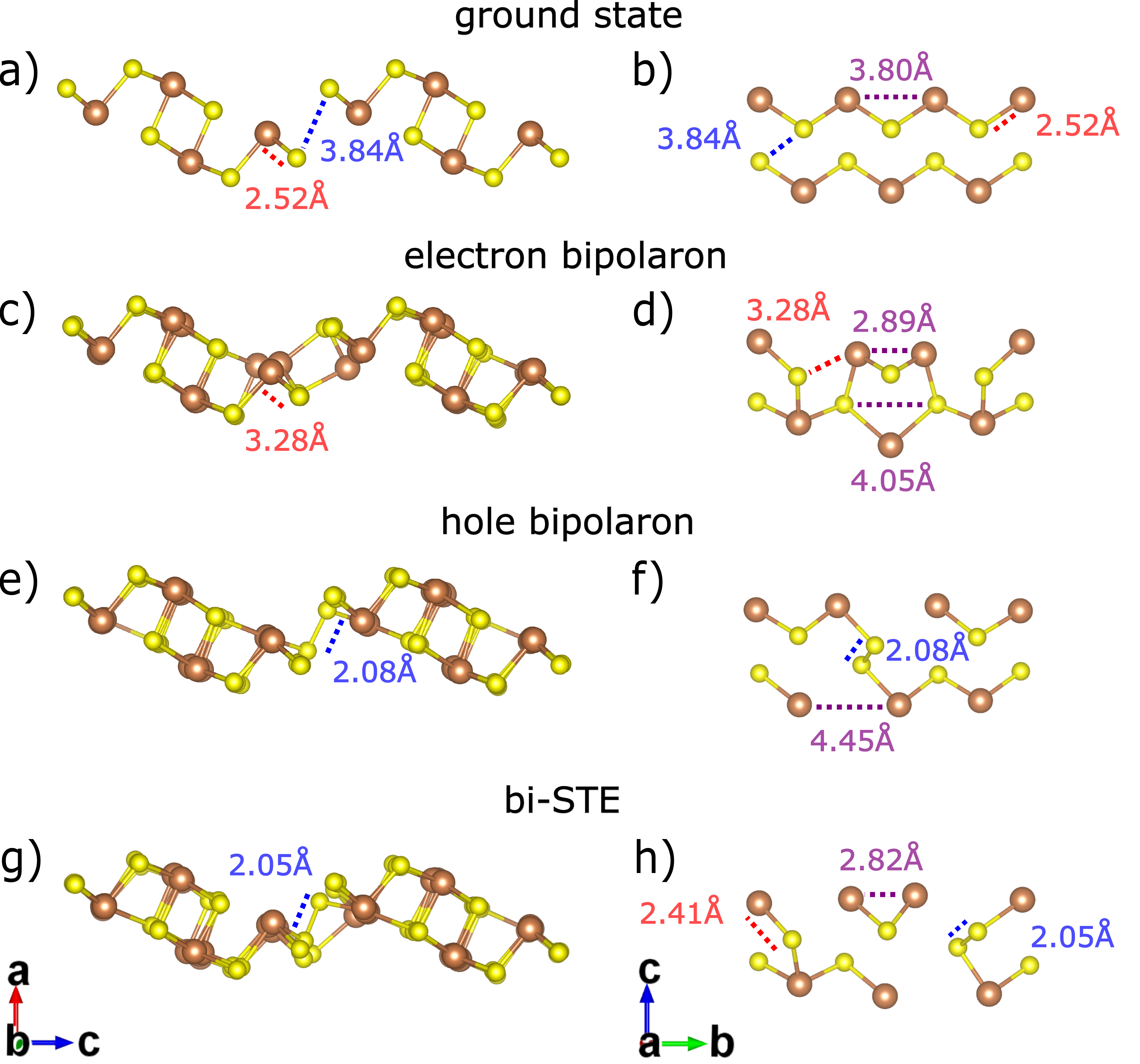}
 \centering
 \caption{Bond distortions arising from excess electrons and holes. (a)(b) The crystal structure and selected bond lengths of  ground state Sb$_2$S$_3$, with characteristic inter-chain S-S (blue), intra-chain Sb-Sb (purple) and Sb-S (red) distances. (c)(d) Distorted structures in the presence of electron bipolaron, with Sb-Sb antimony dimer bond lengths at $2.89$\,\AA.  (e)(f) Distorted structures in the presence of hole bipolaron, with S-S sulphur dimer bond lengths at $2.08$\,\AA. (g)(h) Distorted structures in the presence of bi-STE, with sulphur dimer and antimony bond lengths further reduced to at $2.05$\,\AA\, and $2.82$\,\AA, respectively.}
 \label{fig_bond}
\end{figure}


Spectroscopic studies on Sb$_2$S$_3$ show that the populations of the trapped carriers do not exhibit any saturation at high carrier density of $10^{20}$ \,cm$^{-3}$, which suggests that the trapping is not due to defects as their concentrations are small\cite{yang_ultrafast_2019, https://doi.org/10.1002/advs.202202154}. The polarized PL emission arising from the preferred dipole alignment of the trapped carriers also rule out the role of defects which are typically randomly distributed. These observations support that carriers in antimony dichalcogenides are self-trapped. However, the nature of the carrier trapping is unsolved as a first-principles study found that polarons in these systems exhibit large radii extending over several unit cells with moderate Fr\"{o}hlich coupling constants\cite{doi:10.1021/acsenergylett.2c01464}. Small polarons and therefore carrier self-trapping are unlikely to occur. 

To resolve some of these controversies, we perform a systematic investigation of the polarons and bipolarons using a large $2\times6\times2$ supercell with 480 atoms. To find the polaronic geometry, we apply the bond distortion method to various bonds to break the symmetry of the crystal structure\cite{pham_efficient_2020}, which has shown to give faster convergence by mimicking the experimentally observed polaronic distortions. One excess electron is then added/removed from the supercell and the atomic positions are relaxed (see details in Methods section). We find that there is little resultant structural distortion, with little changes in the wavefunctions of the CBM and VBM states as shown in Figure S2. The excess electron is delocalized over the Sb atoms located on the edges of a [Sb$_4$S$_6$] ribbon, while the excess hole is delocalized over the entire supercell. Due to the delocalized nature of the band edge wavefunction, the formation energy of the excess carrier is calculated as the difference between the total energy of the supercell before and after structural relaxation. The formation energy of 4\,meV is within the error of the DFT method and well below the energy of thermal fluctuations at room temperature, meaning that the formation of a small polaron is unlikely. These results are in agreement with the conclusions reached in the work of Wang \textit{et. al.} \cite{doi:10.1021/acsenergylett.2c01464}.

We next investigate the addition/removal of two excess electrons, and note that the resultant excited carrier density of $1.6 \times 10^{20}$\,cm$^{-3}$ is of the same order of magnitude as the carrier densities used in the transient absorption spectroscopy measurements of Sb$_2$S$_3$\cite{yang_ultrafast_2019}. This motivates us to further explore the possible formation of bipolarons, which have been investigated in conjugated polymers and oxides for their roles on magnetoresistance and transport properties \cite{cohen_small-bipolaron_1984, wagemans_two-site_2008, devreese_frohlich_2009}.

\begin{figure}[t]
 \includegraphics[scale=0.25]{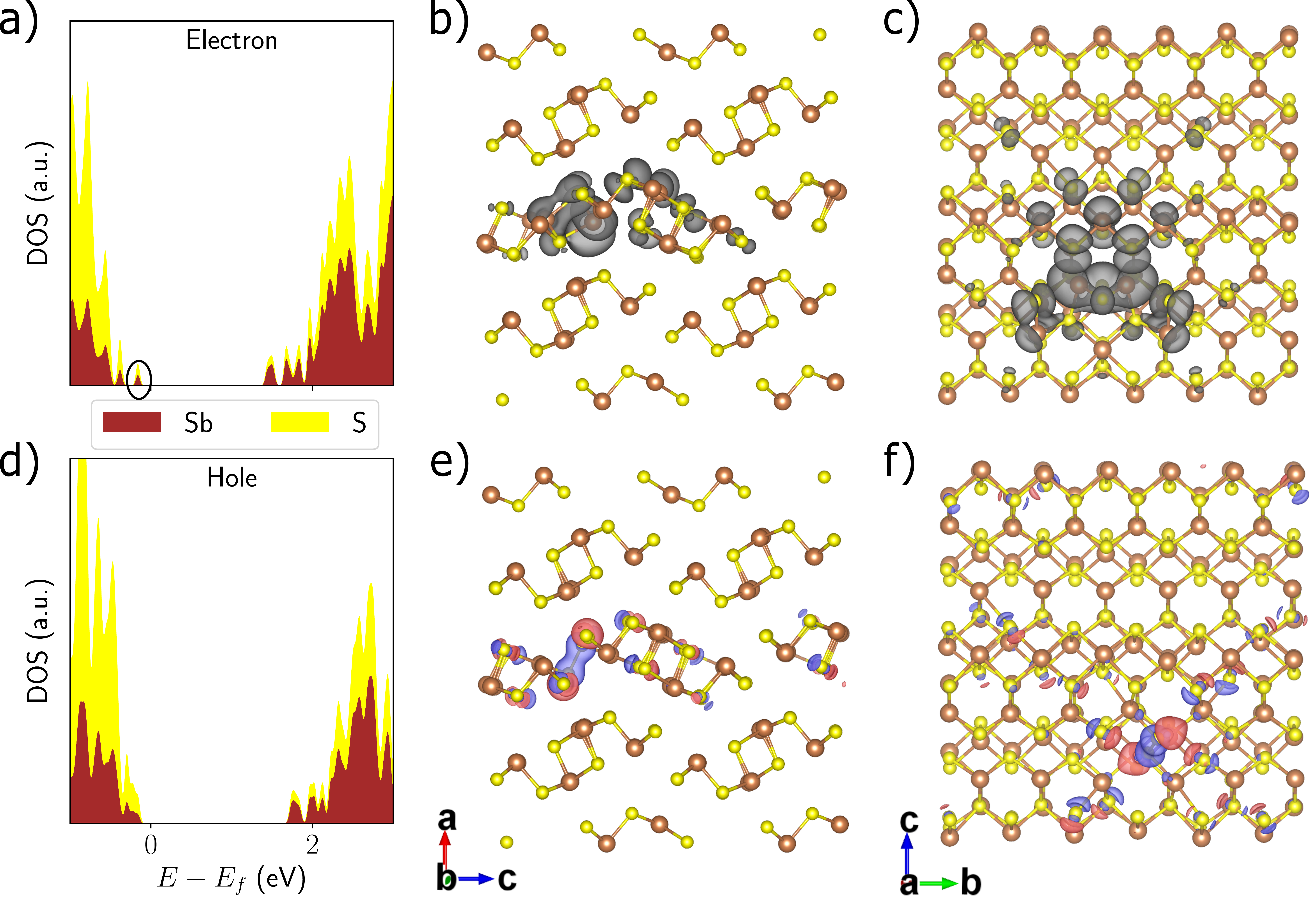}
 \centering
 \caption{Electronic structures of the polaronic states. (a)(d) Projected density of states (PDOS) of the relaxed bipolaronic structures with two excess electrons and holes, respectively. The energetic location of the electron bipolaron and hole bipolaron state are circled. (b)(c) Wavefunction isosurface of the electron bipolaron as viewed from the $b$ and $c$ axis. (e)(f) The difference between the total charge density of the hole bipolaron and the pristine Sb$_2$S$_3$ as viewed from the $b$ and $c$ axis. The blue and red colors represent positive and negative values.}
 \label{fig_bipolaron}
\end{figure}

Similar to the search for polaronic structure, two excess electrons are added/removed from the supercell and the atomic positions relaxed after applying the bond distortion method.
Compared with the ground state geometry (Figure \ref{fig_bond}(a,b)), there is a significant structural distortion with the elongation of the typical Sb-S bond from $2.52$ to $3.28$\,\AA\, with two excess electrons (Figure \ref{fig_bond}(c)). Within the [Sb$_4$S$_6$] chain, antimony dimers form with the nearest Sb-Sb distance decreasing from from $3.80$ to $2.89$\,\AA\,, while the S-S distance increases from $3.80$ to $4.05$\,\AA\, (Figure \ref{fig_bond}(d)). 
Dimer formation in the presence of bipolarons has been observed in oxides such as TiO$_2$, LiNbO$_3$ and BiVO$_4$
\cite{schirmer_electron_2009, chen_double-hole-induced_2014, osterbacka_charge_2022}.
The dimer formation can be understood from energetics: if the energy gained from the formation of the bonding orbitals between the antimony atoms is larger than the Coulomb repulsion between the two excess electrons, the two electron polarons can bind with each other and induce the dimerization\cite{chen_double-hole-induced_2014}.
The electron bipolaron state appears at the top of the valence bands (Figure \ref{fig_bipolaron}(a)), with its wavefunction  localized between two [Sb$_4$S$_6$] chains near the antimony dimers (Figure \ref{fig_bipolaron}(b)(c)). 

Analogously, the introduction of two excess holes results in the formation of a sulphur dimer between two neighbouring [Sb$_4$S$_6$] chains, with the S-S distance decreasing from $3.84$ to $2.08$\,\AA\, (Figure \ref{fig_bond}(e)(f)). 
The formation of the sulphur dimer pushes the state associated with its two holes far into the conduction band, where hybridization with conduction states occurs\cite{chen_double-hole-induced_2014, osterbacka_charge_2022}. For this reason, it is impossible to isolate these two holes for visualization. For illustrative purposes, we plot the difference between the total charge density of the hole bipolaron and the pristine Sb$_2$S$_3$ in Figure \ref{fig_bipolaron}(d-f), resulting in the bonding orbitals between two sulphur atoms in the dimer.
By modifying the defect formation energy expression, we compute the formation energies per carrier for electron and hole bipolarons as $-330$\,meV and $-280$\,meV (Equation \ref{eqn_polaron}).

We then construct a bi-STE from neighbouring electron and hole bipolarons by superimposing their structural distortions within a neutral simulation cell. The ground state electronic configurations are used and the system relaxes to a meta-stable configuration containing both the electron and hole bipolaron. As shown in Figure \ref{fig_bond}(g)(h), both the antimony and sulphur dimers still exist in this bi-STE structure, with their respective bond lengths further reduced compared to the individual bipolaron case. We note that a full description of excitons requires  higher levels of theory than hybrid functionals, but a previous Bethe-Salpeter equation study of Sb$_2$S$_3$ indicates small exciton binding energies of $10-50$\,meV\cite{caruso_excitons_2015}. The formation energy of the bi-STE per exciton is calculated as $-700$\,meV (Equation \ref{eqn_ste}) and larger than the sum of the individual bipolaron, showing that these localized bipolarons further attract each other due to Coulomb interactions. 
The formation energy of the bi-STE is in general agreement with the reported Stokes shift of $600$\,meV in Sb$_2$S$_3$. The slight overestimation from the calculation might arise from the accuracy of the band gap value of HSE06 hybrid functionals, as well as the structural deviation from the true excited state geometry of the bi-STE state.

\section*{Conclusion}

We study electron-phonon coupling in the quasi-1D semiconductor Sb$_2$S$_3$. We first show that the optB86b nonlocal vdw-DF functional is the best choice for reproducing the experimental lattice parameters at reasonable computational cost. The quasi-harmonic approximation is capable of reproducing the thermal expansion coefficient and volumes agree well with experiments. We also find that the absorption edge red shifts by about $200$\,meV from $10$\,K to $300$\,K, a higher value than the corresponding shift observed in most conventional and van der Waals semiconductors. This shows that there is significant electron-phonon coupling in the system. We further investigate the possibility of polaron, bipolaron, and bi-self-trapped-exciton (bi-STE) in the system.
In the presence of one excess carrier, we find that no small polarons are formed in agreement with previous studies.
With two excess carriers per supercell, electron and hole bipolarons can cause the formation of the sulphur and antimony dimers with relatively localized wavefunctions.
The carrier densities of the bipolaron and bi-STE of $1.6 \times 10^{20}$ cm$^{-3}$ in our calculations agree well with experimental values in transient absorption experiments, and the bi-STE formation energy also agrees well with the experimentally measured Stokes shift. 
Our results reconcile some of the conflicting reports on carrier trapping in Sb$_2$S$_3$, and show that intrinsic self trapping can occur without the presence of defect states.
Overall, large electron-phonon coupling and the presence of carrier self-trapping play an important role for the optoelectronic properties of Sb$_2$S$_3$, and might place a fundamental limit on the open circuit voltage of photovoltaic devices and consequently on the maximum efficiency of derived solar cells.

\section*{Methods}
All DFT calculations are performed using the Vienna Ab initio Simulation Package (\texttt{VASP}, v5.4)\cite{kresse_efficient_1996, kresse_efficiency_1996}. The core-valence interaction is described using the projector-augmented wave (PAW) method\cite{blochl_projector_1994}, with 5 valence electrons for Sb ($5s^25p^3$) and 6 valence electrons for S ($3s^23p^4$). The electronic wave functions are expanded in a plane wave basis with an energy cutoff of $400$\,eV, the Brillouin zone is sampled with a $12\times4\times12$ $\Gamma$-centered Monkhorst-Pack\cite{monkhorst_special_1976} $\mathbf{k}$-point grid, and commensurate grids for the supercells. The atoms are relaxed until the Hellman-Feynman force converges below $10^{-2}$\,eV\AA$^{-1}$, and the volume until all components of the stress tensor are below $10^{-2}$\,GPa. 

All phonon dispersions are computed using the finite displacement method with a $2\times6\times2$ supercell containing $480$ atoms ($22.6$\,\AA\,$\times23.2$\,\AA\,$\times22.1$\,\AA) as implemented in the \texttt{Phonopy} package\cite{togo_first_2015}. For phonon dispersions, a non-analytical term is added to the dynamical matrix to treat the long range interaction arising from the macroscopic electric field induced by the polarization of collective ionic motions near $\Gamma$\cite{gonze_dynamical_1997}.

To include the effects of electron-phonon interactions to optical absorption at a given temperature $T$, we evaluate the imaginary part of the frequency-dependent dielectric function $\varepsilon_2(\omega, T)$ within the independent particle approximation using the Williams-Lax theory:
\begin{eqnarray}
\varepsilon_2(\omega, T) = \frac{1}{\mathcal{Z}} \sum_{\mathbf{s}} \braket{\Phi_{\mathbf{s}}({\mathbf{u})}|\varepsilon_2(\omega, T)|\Phi_{\mathbf{s}}({\mathbf{u})}} e^{-E_{\mathbf{s}}/k_\mathrm{B}T},
\label{eqn_band_renorm}
\end{eqnarray}
where $\Phi_{\mathbf{s}}$ is the vibrational wave function in state $\mathbf{s}$ and with energy $E_{\mathbf{s}}$, evaluated within the harmonic approximation, and $\mathcal{Z} = \sum_{\mathbf{s}} e^{-E_{\mathbf{s}}/k_{\mathrm{B}}T}$ is the partition function in which $k_{\mathrm{B}}$ is Boltzmann's constant. For these calculations, we re-compute the phonon frequencies and eigenvectors using non-diagonal supercells\cite{lloyd-williams_lattice_2015}, and then use them as a starting point to evaluate Equation (\ref{eqn_band_renorm}) with Monte Carlo integration accelerated by thermal lines\cite{monserrat_vibrational_2016, monserrat_electron-phonon_2018}.

From the finite temperature dielectric function, the absorption coefficient is given by $\alpha(\omega) = \frac{\omega}{cn(\omega)} \varepsilon_2(\omega)$, where $c$ is the speed of light in vacuum, $\varepsilon_2(\omega)$ is the imaginary part of the dielectric function, and $n(\omega)$ is the real part of the complex refractive index. $n^2(\omega) = \frac{1}{2}(\varepsilon_1 + \sqrt{\varepsilon_2^2+\varepsilon_1^2})$, and where $\varepsilon_1(\omega)$ is the real part of the dielectric function. $\varepsilon_1(\omega)$ is obtained from $\varepsilon_2(\omega)$ through the Kramers-Kronig relation. Convergence tests show that a $2\times6\times2$ supercell and a $2\times2\times2$ electronic $\mathbf{k}$-grid lead to accurate results.

For the fulfillment of the Koopmans' condition, the corrections to the unoccupied Kohn-Sham eigenvalues of the defect-induced single particle levels are calculated as\cite{chen2013correspondence}
\begin{eqnarray}
   \epsilon_{\mathrm{corr}}^{\mathrm{KS}} = \frac{-2}{q}E_{\mathrm{corr}}
\end{eqnarray}
where $q$ is the charge of the defect, and $E_{\mathrm{corr}}$ is the finite size electrostatic correction. $E_{\mathrm{corr}}$ is computed using \texttt{sxdefectalign}\cite{PhysRevLett.102.016402} with an anisotropic screening where the diagonal terms of the high-frequency dielectric tensor are $\varepsilon_{xx}=11$, $\varepsilon_{yy}=8$, $\varepsilon_{zz}=12$\cite{wang_lone_2022}.

For polaron and self-trapped exciton (STE) calculations, we use a $2\times6\times2$ supercell to minimize spurious interactions between periodic images, and sample only the $\Gamma$-point. After the addition or removal of an electron from the supercell, structural relaxation is performed using spin-polarized calculations whereby the supercell lattice parameters are fixed and the atoms allowed to move, with the same force convergence criterion of $10^{-2}$\,eV\AA$^{-1}$. The binding energy of the electron or hole bipolaron per electron ($E_p$) can be estimated using the following formula for defect formation energy calculations\cite{freysoldt_first-principles_2014}:
\begin{eqnarray}
  E_p = \frac{E_q[\mathrm{bipolaron}] - E[\mathrm{pristine}] + 2qE_\mathrm{edge} + E_{\mathrm{corr}}}{2}
\label{eqn_polaron}
\end{eqnarray}
where $E_q[\mathrm{bipolaron}]$ is the total energy of the distorted supercell of the bipolaronic state, $E$[pristine] is the total energy for the perfect crystal using an equivalent supercell, with $q$ denoting the excess of charge of an electron or hole, and $E_{\mathrm{edge}}$ is the energetic position of the CBM or VBM. The diagonal components of the static dielectric tensor are used with $\varepsilon_{xx}=94$, $\varepsilon_{yy}=13$, $\varepsilon_{zz}=99$\cite{wang_lone_2022}.

The formation energy of the bi-STE per exciton ($E_s$) is calculated as 
\begin{eqnarray}
  E_s = \frac{E[\text{bi-STE}] - E[\mathrm{pristine}] -2E_g}{2}
\label{eqn_ste}
\end{eqnarray}
where $E[\text{bi-STE}]$ is the total energy of the distorted supercell of the bi-STE state with the ground state electronic configuration.


Crystal structures and isosurfaces are visualized using \texttt{VESTA}\cite{momma_vesta_2008} and graphs are plotted by \texttt{sumo}\cite{m_ganose_sumo_2018} and custom scripts.

\section*{Acknowledgements}
Y.L acknowledges funding support from the Simons Foundation (Grant 601946) and A*STAR under its Young Achiever Award.
J.W. acknowledges funding from the Swedish Research Council (2019-03993). B.M. acknowledges support from a UKRI Future Leaders Fellowship (Grant No. MR/V023926/1), from the Gianna Angelopoulos Programme for Science, Technology, and Innovation, and from the Winton Programme for the Physics of Sustainability. This work was performed using resources provided by the Cambridge Service for Data Driven Discovery (CSD3) operated by the University of Cambridge Research Computing Service (www.csd3.cam.ac.uk), provided by Dell EMC and Intel using Tier-2 funding from the Engineering and Physical Sciences Research Council (capital grant EP/T022159/1), and DiRAC funding from the Science and Technology Facilities Council (www.dirac.ac.uk). We are also grateful for computational support from the UK national high performance computing service, ARCHER, for which access was obtained via the UKCP consortium and funded by EPSRC grant EP/P022561/1. 
We also acknowledge computational resources provided by the Swedish National Infrastructure for Computing (SNIC) at C3SE and PDC.


\bibliography{sb2s3}

\end{document}